\documentclass[longauth]{aa}  

\usepackage{graphicx}
\usepackage{txfonts}
\usepackage{natbib}
\bibpunct{(}{)}{;}{a}{}{,} 
\renewcommand{\vec}{\boldsymbol}
\usepackage{siunitx}
\usepackage{xcolor}
\usepackage{multirow}

\begin{document} 

   \title{Gaia21blx: Complete resolution of a binary microlensing event in the Galactic disk}
   
   \subtitle{}
   
     \author{P. Rota \inst{\ref{Salerno},\ref{INFN}}
    \and
    V. Bozza \inst{\ref{Salerno},\ref{INFN}}
    \and 
    M. Hundertmark \inst{\ref{Heidelberg}} 
    \\
    and
    \\
    E. Bachelet \inst{\ref{ipac}}
    \and
    R. Street \inst{\ref{LCO}}
    \and
    Y. Tsapras \inst{\ref{Heidelberg}}
    \and
    A. Cassan \inst{\ref{inst:iap}}
    \and 
    M. Dominik \inst{\ref{St.Andrews}}
    \and
    R. Figuera Jaimes\inst{\ref{inst:mas}, \ref{inst:puc}}
    \and 
    K. A. Rybicki\inst{\ref{WIS}}
    \and
    J. Wambsganss \inst{\ref{Heidelberg}}
    \and
    {\L}. Wyrzykowski\inst{\ref{OAWarsaw}}
    \and
    P. Zieli\'{n}ski\inst{\ref{NCU}}
    \\
    The OMEGA Key Project
    \\
    and
    \\
    M. Bonavita \inst{\ref{UERO}}
    \and
    T. C. Hinse \inst{\ref{SDU}}
    \and
    U. G. J{\o}rgensen \inst{\ref{NBI}}
    \and
    E. Khalouei \inst{\ref{RIBS}}
    \and
    H. Korhonen \inst{\ref{NBI}}
    \and
    P. Longa-Pe{\~n}a \inst{\ref{UA}}
    \and
    N. Peixinho \inst{\ref{IACE}}
    \and
    S. Rahvar \inst{\ref{SUT}}
    \and
    S. Sajadian \inst{\ref{IUT}}
    \and
    J. Skottfelt \inst{\ref{CEI}}
    \and
    C. Snodgrass \inst{\ref{UERO}}
    \and 
    J. Tregolan-Reed \inst{\ref{IIACP}}
    \\ 
    The MiNDSTEp consortium
    }
    \institute{Dipartimento di Fisica E.R. Caianiello, Università di Salerno, Via Giovanni Paolo II 132, I-84084, Fisciano, Italy\label{Salerno}
    \and
    Istituto Nazionale di Fisica Nucleare, Sezione di Napoli, I-80126, Napoli, Italy\label{INFN}
    \and
    Astronomisches Rechen-Institut, Zentrum f{\"u}r Astronomie der Universit{\"a}t Heidelberg (ZAH), 69120 Heidelberg, Germany \label{Heidelberg}
    \and
    Las Cumbres Observatory, 6740 Cortona Drive, Suite 102,93117 Goleta, CA, USA \label{LCO}
    \and
    University of St Andrews, Centre for Exoplanet Science, School of Physics \& Astronomy, North Haugh, St Andrews, KY16 9SS, United Kingdom \label{St.Andrews}
    \and
    Millennium Institute of Astrophysics MAS, Nuncio Monsenor Sotero Sanz 100, Of. 104, Providencia, Santiago, Chile \label{inst:mas}
    \and 
    Instituto de Astrof\'isica, Facultad de F\'isica, Pontificia Universidad Cat\'olica de Chile, Av. Vicu\~na Mackenna 4860, 7820436 Macul, Santiago, Chile \label{inst:puc}
    \and
    Institut d'Astrophysique de Paris, Sorbonne Universit\'e, CNRS, UMR 7095, 98 bis bd Arago, F-75014 Paris, France
    \label{inst:iap}
    \and IPAC, Mail Code 100-22, Caltech, 1200 E. California Blvd., Pasadena, CA 91125 USA \label{ipac}
    \and
    Astronomical Observatory, University of Warsaw, Al.~Ujazdowskie~4, 00-478~Warszawa, Poland\label{OAWarsaw}
    \and
    Department of Particle Physics and Astrophysics, Weizmann Institute of Science, Rehovot 76100, Israel \label{WIS}
    \and
    Institute of Astronomy, Faculty of Physics, Astronomy and Informatics, Nicolaus Copernicus University in Toru{\'n}, Grudzi\k{a}dzka 5, 87-100 Toru{\'n}, Poland \label{NCU}
    \and
    Centre for ExoLife Sciences, Niels Bohr Institute, University of Copenhagen, {\O}ster Voldgade 5, 1350 Copenhagen, Denmark \label{NBI}
    \and
    Department of Physics, Isfahan University of Technology, Isfahan 84156-83111, Iran  \label{IUT}
    \and
    Astronomy Research Center, Research Institute of Basic Sciences, Seoul National University,
    1 Gwanak-ro, Gwanak-gu, Seoul 08826, Korea \label{RIBS}
    \and
    Instituto de Astrofísica e Ciências do Espaço , Universidade de Coimbra , 3040-004 Coimbra,Portugal \label{IACE}
    \and
    Unidad de Astronom{\'{\i}}a, Universidad de Antofagasta, Av.\ Angamos 601, Antofagasta, Chile \label{UA}
    \and
    Department of Physics, Sharif University of Technology, PO Box 11155-9161 Tehran, Iran \label{SUT}
    \and
    Instituto de Investigación en Astronomia y Ciencias Planetarias, Universidad de Atacama, Copiapó, Atacama, Chile \label{IIACP}
    \and
    Centre for Electronic Imaging, Department of Physical Sciences, The Open University, Milton Keynes, MK7 6AA, UK \label{CEI}
    \and
    Institute for Astronomy, University of Edinburgh, Royal Observatory,Edinburgh EH9 3HJ, UK \label{UERO}
    \and
    University of Southern Denmark, Department of Physics, Chemistry and Pharmacy, SDU-Galaxy, Campusvej 55, 5230 Odense M, Denmark \label{SDU}
    }
    \date{Received ??; accepted ??}
  \abstract
   {Gravitational microlensing is a method that is used to discover planet-hosting systems at distances of several kiloparsec in the Galactic disk and bulge. We present the analysis of a  microlensing event reported by the Gaia photometric alert team that might have a bright lens.} 
   {In order to infer the mass and distance to the lensing system, the parallax measurement at the position of Gaia21blx was used. In this particular case, the source and the lens have comparable magnitudes and we cannot attribute the parallax measured by Gaia to the lens or source alone.}
   {Since the blending flux is important, we assumed that the Gaia parallax is the flux-weighted average of the parallaxes of the lens and source. Combining this assumption with the information from the microlensing models and the finite source effects we were able to resolve all degeneracies and thus obtained the mass, distance, luminosities and projected kinematics of the binary lens and the source.}
   { According to the best model, the lens is a binary system at $2.18 \pm 0.07$ kpc from Earth. It is composed of a G star with $0.95\pm 0.17\,M_{\odot}$ and a K star with $0.53 \pm 0.07 \, M_{\odot}$. The source is likely to be an F subgiant star at $2.38 \pm 1.71$ kpc with a mass of $1.10 \pm 0.18 \, M_{\odot}$. Both lenses and the source follow the kinematics of the thin-disk population. We also discuss alternative models, that are disfavored by the data or by prior expectations, however.}
   {}

   \maketitle
   
\section{Introduction}
Gravitational microlensing occurs when the light of the background source star is deflected by a foreground object (called lens). The result is a time-variable magnification of the flux signal. By studying the shape of the light curve, we can determine the properties of the lens \citep{Mao2012, Gaudi2012, Tsapras2018}. This phenomenon was first hypothesized by \citet{Einstein1936}, and \citet{Paczynski1986} described how the method might be applied to find dark matter in the Galactic halo. 
In recent years, microlensing has become a unique tool for the identification of planets around faint objects. These planets cannot be detected with other techniques.
With gravitational microlensing, we can in particular study binary systems regardless of the luminosity of their components, because the result of their gravitational influence on the light of a background source is detected, not their intrinsic luminosity. In this way, we are able to find exoplanets orbiting single or binary stars \citep{Gould1992, Bond2004, Bennett2016}, brown dwarfs \citep{Bozza2012, Ranc2015, Herald2022}, or compact objects such as black holes or stellar remnants \citep{Shvartzvald2015, Wyrzykowski2016, Sahu2022, Lam2022, Mroz2022}.

Numerous surveys have been carried out by various collaborations such as Optical Gravitational Lensing Experiment (OGLE; \citealt{Udalski2003}), Microlensing Observations in Astrophysics (MOA; \citealt{Bond2001,Sumi2003}), and Korea Microlensing Telescope Network (KMTNet; \citealt{Kim2016}) as well as follow-up surveys such as Microlensing Follow-Up Network ($\mu$Fun; \citealt{Gould2008}), Probing Lensing Anomalies Network collaboration (PLANET; \citealt{Albrow1998}), RoboNet \citep{Tsapras2009}, Microlensing Network for the Detection of Small Terrestrial Exoplanets (MiNDSTEp; \citealt{Dominik2010}), and the Observing Microlensing Events of the Galaxy Automatically Key Project (OMEGA Key Project; Bachelet et al. in prep.). \\
With the new generation of large-sky surveys, 
we are able to find microlensing events in the entire sky. For this purpose, the work of Gaia is of fundamental importance given its mission to build an extremely detailed three-dimensional map of the Milky Way \citep{Gaia2016}. 
It is worth noting that  microlensing discoveries have historically been confined to a small region in the Bulge, in contrast to Gaia’s probing of the whole galaxy. 
Gaia's early-alert capabilities \citep{Hodgkin2021} played a crucial role in obtaining timely follow-up observations for many important microlensing events discovered by the mission \citep{Gaia2016}. More than 350 microlensing events were detected \citep{Wyrzykowski2023} and over 1700 were predicted using astrometric simulations \citep{Kluter2022}. This is important because most of the observed microlensing events occur in the Galactic disk, where the duration of the event is longer, such that high-order effects can be detected. \\
Several events have been studied to date with interesting results, that we introduce below. 
For example, the single lens event Gaia18cbf \citep{Kruszynska2022} has a very long timescale of $\sim$ 490~days, in which the photometric data revealed a lens of $\sim$ 2~$M_{\odot}$. We also mention Gaia19bld \citep{Rybicki2022, Bachelet2022, Cassan2022}, which is a high-magnification event with a dark lens of~1.1~$M_{\odot}$. This type of events occurs when the source comes very close to the central caustic \citep{Griest1998}, which is the set of points where the magnification diverges. Gaia has also discovered binary lens events. The spectacular event Gaia16aye is most notable \citep{Wyrzykowski2020}. It is one of the first microlensing events detected by Gaia. Its timescale is so long (over $\sim$ 500 days) that it was possible to obtain a full orbital solution for this system.

The most crucial advantage of Gaia observations of microlensing events is the fact that Gaia also obtains astrometric time series with submilliarcsecond precision \citep{Rybicki2018}.
With the data to be released in Gaia DR4, it will therefore be possible for many of the sources observed between 2014 and 2019 to obtain the size of the angular Einstein ring radius $\theta_\mathrm{E}$, and based on this, to obtain mass-distance relation for the lenses, in particular, for dark lenses \citep{Dominik2000, Belokurov2002, McGill2018, Kluter2022, Jablonska2022}.

The DR3 Gaia data release covered 34 months of observations in which every target was observed about 40~times \citep{Gaia2016}. With these measurements, it was possible to obtain various parameters such as the parallax, the proper motion, and other quantities that are useful for studying the microlensing event. 
Difficulties arise when the parallax reported by Gaia corresponds to
a microlensing event in which the blend flux (the flux of nearby unresolved objects that are not affected by microlensing) is very high and mainly generated by the lens. A possible interpretation is that the source and the lens both contribute to the total flux, making the parallax measured by Gaia (and also the proper motion) a flux-weighted average of the parallaxes of lens and source. With this assumption, it is possible to derive a constraint on the lens parameters that will then be explained in detail. 
This is what we have done in the study of the microlensing event Gaia21blx, which also benefits from follow-up by the OMEGA collaboration. The paper is organized as follows. Section 2 is dedicated to the observations and data reduction. In Section 3, we introduce the modeling of this event, which leads to four degenerate solutions. We discuss the means to break this degeneracy in Section 4, and we propose a method for determining the most likely physical model after an accurate evaluation of the probabilities of possible alternatives, and we derive the physical parameters of the lens.
In Section 5, we discuss the kinematics of the lens and source. In the last section, we summarize the results we obtained.
   

\section{Observations}

Gaia21blx is located in the Galactic disk at (R.A., decl.) = $(14^h53^m15^s.42, -62^{\circ}01'30''.61)$,  corresponding to Galactic coordinates $l=316.69911^{\circ}$, $b=-2.45443^{\circ}$. The alert was published on 22 March 2021 by the Gaia science alerts (GSA) system \citep{Hodgkin2021} with a magnitude of $\sim 16.64$. The photometric measurements obtained by Gaia consist of a wide G band \citep{Jordi2010} performed on a monthly period. This is publicly available at the GSA. The photometric errors for the 153 measurements used in this work were calculated following the procedure described by \citet{Kruszynska2022}.  

In response to the Gaia alert, high-cadence follow-up observations were obtained by the OMEGA Project, using the Las Cumbres Observatory Global Telescope Network (LCO) of 1m telescopes \citep{Brown2013}. Data were collected in SDSS-g' and -i' bandpasses using the LCO Sinistro instruments, at LCO sites at the Cerro Tololo Interamerican Observatory, Chile (CTIO), the South African Astronomical Observatory (SAAO), Sutherland, South Africa, and the Siding Spring Observatory in Australia.  The instrumental signatures were removed from the resulting image data using LCO Beautiful Algorithms to Normalize Zillions of Astronomical Images pipeline (BANZAI; \citealt{McCully2018}), after which difference image analysis (DIA) was used to derive time-series photometry using the pyDANDIA software package \citep{Bramich2008,Bramich2013}. The process of calibrating the reference images in the g' and i' bands is explained in \citet{street2024}. In short, the instrumental magnitudes are aligned to the VST Photometric H$\alpha$ Survey of the Southern Galactic Plane and Bulge (VPHAS; \citealt{Drew2014}) or the Gaia synthetic photometry \citep{Montegriffo2023}, depending on the availability of the catalog for this part of the sky, after an astrometric crossmatch. The linear relation obtained is ultimately propagated to the entire photometry, as are the errors. The Sinistro instruments on the LCO 1m network are designed to be extremely similar. Their detectors have very similar read-noise properties, and the difference between the read noise of every detector is negligible in the final photometry. The LCO BANZAI pipeline also removes the instrument-specific or telescope optics signature using calibration frames specific to each camera. The image data from different telescopes are only combined after this point in order to process the instrument-signature-corrected images with the DIA pipeline. Processing all images using a single reference image is advantageous at this stage because DIA inherently computes the time-variable component of the flux relative to that image (in each bandpass). If the data are separated by instrument, then the light curve from each instrument is calculated relative to reference images taken under different atmospheric conditions, meaning that in order to combine the datasets for analysis, a second step of calibrating the photometry between the different reference images is required. This introduces an additional source of uncertainty. When the data are combined into a single dataset, this additional step is not required. This procedure was also adopted in previous works and has become a standard for OMEGA data \citep{Rybicki2022,Olmschenk_2023}. 

A fourth dataset is obtained by the MiNDSTEp collaboration \citep{Dominik2010} using the Danish 1.54 Telescope located at ESO La Silla in Chile. This telescope is equipped with a multiband Electron Multiplying CCD Camera instrument (EMCCD; \citealt{Skottfelt2015}), and the images used were reduced using PyDandia. In total we collected 153 data points in the Gaia G band that were mostly taken before the event, and we have 196 data points in i$'$ band and 89 data points in g$'$ band. Following common practice in microlensing \citep{Yee2012, Miyake2012}, we rescaled the error bars following $\sigma_i^2=k\sqrt{\sigma_{i,\mathrm{orig}}^2+e_\mathrm{min}^2}$, where in this case,  we adopted $e_\mathrm{min}=0.03$ for the LCO telescope in i' band, which covers the peak of the event. We set $e_\mathrm{min}=0$ for the other telescopes, and $k$ ensures that $\chi^2/$d.o.f.=1 for the best model.

\begin{table*}
\centering
\small
\caption{Data from telescopes with their corresponding rescaling parameters and the limb-darkening coefficients as derived from stellar models (see Sect. 4.2)}
\label{tab:data}
\begin{tabular}{cccccc}
       Telescope  & Band & N. data& k & $e_{min} $&Limb darkening \\
       \hline
        LCO$_\mathrm{gp}$ & SDSS-g' & 85 &1.275 & 0&$0.642$\\
        LCO$_\mathrm{ip}$ & SDSS-i' & 223 &0.727 & 0.03&$0.415$\\
         Danish 1.54m & R & 28 & 2.846 & 0& $0.461$\\
        Gaia & G-Gaia& 153 & 1.039 & 0&$0.514$\\
         \hline
\end{tabular}  
\end{table*}

\section{Modeling}
The microlensing effect can be described by the Einstein angle, which describes the angular scale of the event,
\begin{equation}
    \theta_\mathrm{E}=\sqrt{\frac{4G M \pi_\mathrm{rel}}{c^2}}
\end{equation}
where $G$ is the gravitational constant, $c$ is the vacuum speed of light, $M$ is the total lens mass, and $\pi_\mathrm{rel}=\mathrm{au}/D_\mathrm{L}-\mathrm{au}/D_\mathrm{S}$ is the relative source-lens parallax, with $D_\mathrm{L}$ the lens distance and $D_\mathrm{S}$ the source distance. The simplest case of a microlensing event, the  point-source single lens, can be described by three parameters: the time of the closest approach between lens and source $t_0$, the impact parameter in units of Einstein angle $u_0$, where the source flux reaches maximum, and the Einstein time $t_\mathrm{E}=\theta_\mathrm{E}/\mu$, where $\mu$ is the relative proper motion in the geocentric reference frame between lens and source. In addition to these is the finite source effect $\rho_*=\theta_*/\theta_\mathrm{E}$, with $\theta_*$ the angular radius of the source. The inclusion of a second lens requires the addition of three more parameters: the separation in units of Einstein angle between the two lenses $s$, the mass ratio $q$, and the angle between the source trajectory and the orthogonally projected separation vector from the secondary lens to the primary lens $\alpha$.
Based on the values of $s$ and $q$, three different configurations are possible: the close configuration, where $s<1$, the intermediate configuration where $s \sim 1$, and the wide configuration where $s>1$ \citep{Dominik1999}. With this configuration, we have the static binary lens model where we approximate the relative lens-source motion as rectilinear, assuming a short timescale. For long timescales, high-order effects often need to be included. The first effect is the annual parallax effect, which is caused by the motion of the Earth around the Sun. This effect on  the microlensing light curves is quantified by \citep{Gould2000} 

\begin{equation}
    \pi_\mathrm{E}=\frac{\pi_\mathrm{rel}}{\theta_\mathrm{E}}\,
\end{equation}
and it depends on the direction of the proper motion in the sky. The parallax vector is defined as $\vec{\pi}_\mathrm{E} = \pi_\mathrm{E}\,\hat{\mu}_{LS}$ \citep{An2002, Gould2004} and allows us to analyze this effect using the two parameters $\pi_{\mathrm{E},\mathrm{N}}$ and $\pi_{\mathrm{E},\mathrm{E}}$, which are its northern and eastern component.
Another way to determine the parameters of the system is to consider the satellite parallax \citep{Refsdal1966, Gould1992, Gould2009}. Observing the same event simultaneously from two different locations (Earth and space) telescopes measure a different $u_0$ and $t_0$. For observations from a satellite, a fourfold degeneracy arises, because the sign of $u_0$ cannot be determined. This degeneracy can be broken by measuring orbital motion, or if  this cannot be detected, it can be reduced to a twofold degeneracy that can be broken with the Rich argument \citep{CalchiNovati2015}.
The second effect is the orbital motion of the two lenses around their common center of mass. In principle, it would be preferable to use the full Keplerian orbit parameterization \citep{Skowron2011}. 
There is no reason to assume zero eccentricity for a stellar binary system a priori. However, the introduction of too many unconstrained parameters would complicate the fit without any benefits. A minimum fit for orbital motion is typically made considering two components, but this leads to unphysical orbital trajectories \citep{Bozza2021}. We preferred to opt for a fit that included the three components of the angular velocity of the system, assuming a circular orbit \citep{Skowron2011, Bozza2021}, to explore physical orbital trajectories. We show below that even these additional parameters are poorly constrained, and make a full Keplerian fit superfluous. 
We then describe the orbital motion with the following parameters: the parallel component of the projected angular velocity parallel to the lens axis at time $t_0$, $\gamma_{\parallel}=(\mathrm{d}s/\mathrm{d}t)/s$, the perpendicular component $\gamma_{\perp}=-(\mathrm{d}\alpha/\mathrm{d}t)$, and the component of the angular velocity along the line of sight $\gamma_z=(\mathrm{d}s_z/\mathrm{d}t)/s$, where $s_z$ is the separation in units of Einstein radius between the two lenses along the line of sight.

The parameters presented above are necessary to describe the model flux $F(t)$ of the magnified source as a function of time,
\begin{equation}
    F(t)=A(t)F_\mathrm{S}+F_\mathrm{b}
\end{equation}
where $A(t)$ is the magnification of the source flux as a function of time, $F_\mathrm{S}$ is the baseline of the source flux, and $F_\mathrm{b}$ is the blend flux, or the flux that is not affected by microlensing.

To explore the parameter space, we used the Real-Time Microlensing Modelling platform \footnote{https://www.fisica.unisa.it/GravitationAstrophysics/RTModel.htm} (RTModel) based on VBBinaryLensing codes \citep{Bozza2010, Bozza2018, Bozza2021}. RTModel fully explores all corners of the parameter space of binary lensing based on a complete library of templates \citep{MaoDiStefano1995,Liebig2015} that covers all possible classes of light curves that arise from different caustic topologies and different trajectories of the source. RTModel returns two main models: a close binary and a wide binary model, whose parameters are shown in Table \ref{tab:21blxmodbin}, with the source trajectory orthogonal to the binary lens axis and crossing the cusp of the central caustic between the two lenses. In both cases, the fit significantly improved when we included the annual parallax. In this case, the reflection symmetry around the lens axis is broken, and we must distinguish models with positive and negative $u_0$, which would be equivalent in the static binary case (see Table \ref{tab:21blxmodpar}). We labeled the four models as C+, C-, W+, and W-, where C or W distinguish the close and wide binary, respectively, and the sign refers to the sign of the impact parameter $u_0$. In addition to the parallax, we also searched for orbital motion and obtained a modest improvement. Nevertheless, the inclusion of orbital motion affects the error bars of the parallax components and should therefore be considered even when the orbital motion remains poorly constrained. The physical consequences of the orbital motion were therefore not considered because the parameters were not robustly determined. From the best solution obtained with RTModel, we ran Markov chains to completely explore the local $\chi^2$ minima to obtain full information about the uncertainties and the correlations in the parameters. We also considered the possibility of a binary source single-lens system but this gave a $\chi^2 \approx 730$, which is very far from the chi-square values obtained by the best models ($\Delta \chi^2 \approx 250$). We therefore excluded this configuration from the analysis.

\begin{figure}[h]
\centering
\includegraphics[width=0.44\textwidth]{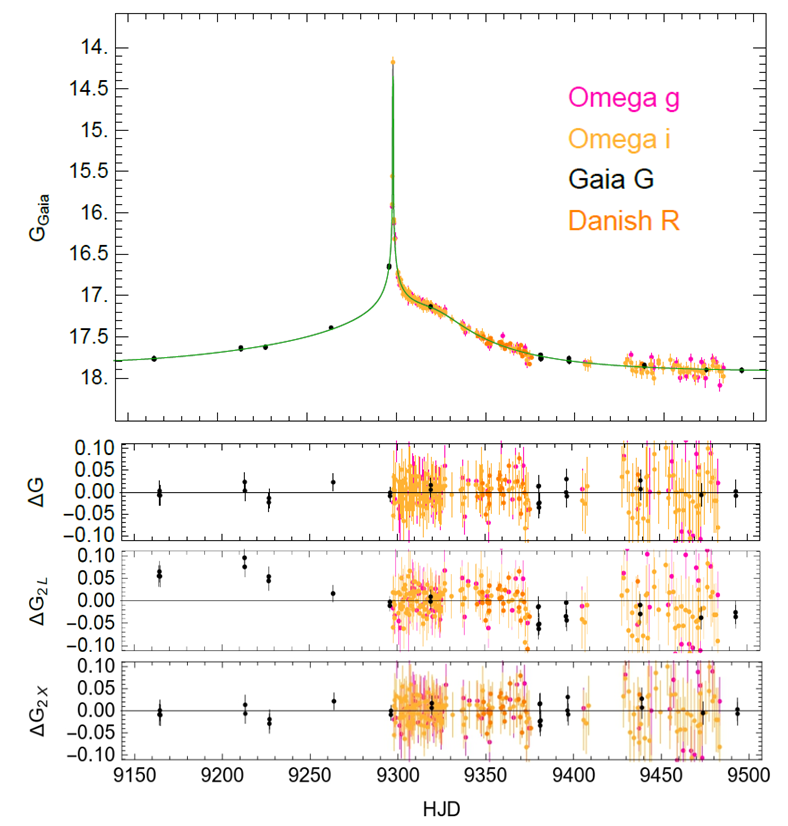}
    \caption{Light curve with residuals of the best model, C+, for the microlensing event Gaia21blx. Below the residuals of the best model we also include the residuals for the static configuration (labeled 2L) and the binary lensing model with only the parallax and no orbital motion (labeled 2X).}
    \label{fig:gd21blxlc}
\end{figure}

\begin{figure}[h]
\centering
\includegraphics[width=0.5\textwidth]{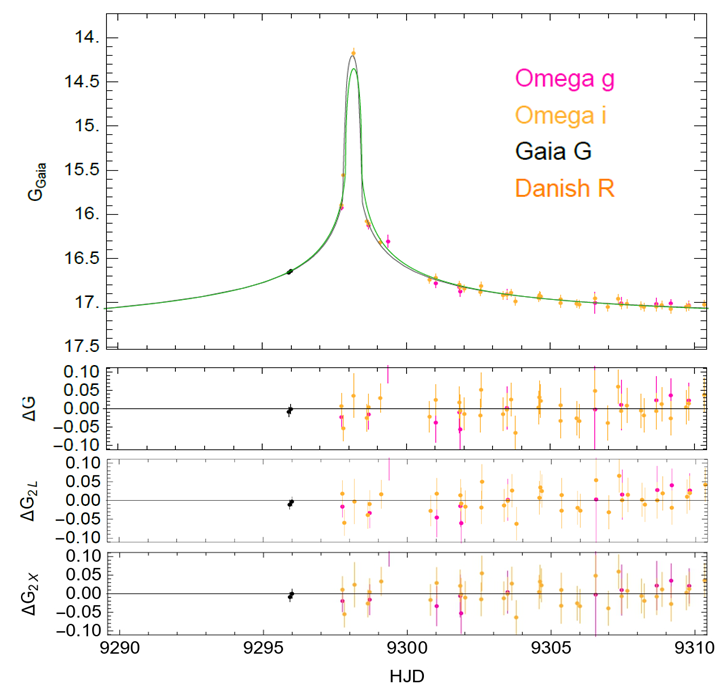}
    \caption{Zoom of the light curve with residuals of the best model, C+, at the peak. The gray light curve is the microlensing light curve as seen from Earth, and the green line is the light curve as seen from Gaia. Below the residuals of the best model, we also include the residuals for the static configuration (labeled 2L) and the binary lensing model with only the parallax and no orbital motion (labeled 2X).}
    \label{fig:gd21blxlcz}
\end{figure}

\begin{figure}[!h]
\centering
\includegraphics[width=0.5\textwidth]{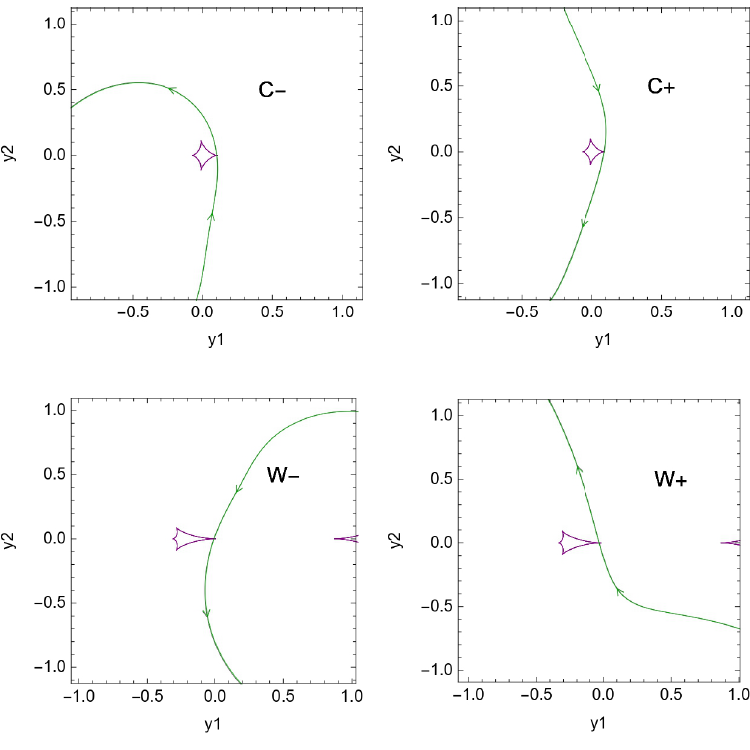}
    \caption{Caustic configuration (in purple) for all the models with the source trajectories (green lines with arrows). At the top lies C- on the left, and on the right lies C+. At the bottom on the left lies the W- model, and the W+ model is shown on the right. The coordinates are centered with respect to the center of mass of the two lenses and are in units of $\theta_\mathrm{E}$}
    \label{fig:gd21blxcau}
\end{figure}

In Figure \ref{fig:gd21blxlc} we show the C+ model (the zoom of the peak is shown in Fig. \ref{fig:gd21blxlcz}). The other models are indistinguishable. All the caustic configurations and the source trajectories are shown in Fig. \ref{fig:gd21blxcau}. The final parameters, including the orbital motion, are shown in Table \ref{tab:21blxmod}. We can easily note that the $\chi^2$ values are similar. At first, this forced us to retain all competitive models. 

\begin{table*}
\centering
\small
\caption{Parameters of the two binary lens models obtained with RTModel after a Markov chain. The parameters are obtained combining ground-based data and Gaia data.} 
\label{tab:21blxmodbin}
\begin{tabular}{cccc}
\hline
Parameters & (Unit) & Close & Wide\\
\hline
$t_\mathrm{E}$ & days &  $167.4^{+17.7}_{-3.2}$  & $79.2_{-1.2}^{+2.4}$ \\  [4pt] 
$t_0$ & HJD-2450000 &  $9299.05^{+0.21}_{-0.07}$ & $9299.34^{+0.20}_{-0.20}$ \\  [4pt]
$u_0$ & & $0.0560^{+0.0022}_{-0.0035}$  & $0.1062^{+0.0175}_{-0.0175}$ \\ [4pt]
$\rho_{*}$ & $10^{-3}$& $1.69^{+0.05}_{-0.19}$   & $3.52^{+0.16}_{-0.13}$ \\ [4pt]
$\alpha$ & &   $-4.809^{+0.007}_{-0.020}$  & $-4.856^{+0.003}_{-0.004}$ \\ [4pt]
$s$ & & $0.364^{+0.003}_{-0.014}$   & $1.867^{+0.017}_{0.017}$	\\ [4pt]
$q$ & &  $0.876^{+0.049}_{-0.146}$   & $0.176^{+0.009}_{-0.009}$ \\ [4pt]
\hline    
$\chi^2$ &  & 598.3 & 657.5 \\   
\end{tabular}
\end{table*}

\begin{table*}
\centering
\small
\caption{Parameters of the four cases of the binary lensing model with the parallax obtained from modeling the light curves using the Markov chain. The parameters are obtained by combining ground-based and Gaia data.} 
\label{tab:21blxmodpar}
\begin{tabular}{cccccc}
\hline
Parameters & (Unit) & C+ & C- & W+ & W-\\
\hline
$t_\mathrm{E}$ & days & $124.4^{+9.9}_{-4.4}$ & $120.0^{+11.0}_{12.0}$ & $90.7_{-7.6}^{+6.4}$ & $97.2^{+3.4}_{-2.2}$ \\  [4pt] 
$t_0$ & HJD-2450000 & $9301.31^{+0.39}_{-0.30}$  & $9300.70^{+0.29}_{-0.24}$ &$9297.34^{+0.08}_{-0.27}$ &$9297.73^{+0.31}_{-0.31}$ \\  [4pt]
$u_0$ & & $0.1003^{+0.0052}_{-0.0081}$ & $-0.0954^{+0.0086}_{-0.0065}$ &$0.0365^{+0.0100}_{-0.0034}$  &$-0.0175^{+0.0137}_{-0.0137}$  \\ [4pt]
$\rho_{*}$ & $10^{-3}$& $2.28^{+0.10}_{-0.19}$  & $2.38^{+0.26}_{-0.26}$ & $3.12^{+0.28}_{-0.24}$ & $2.86^{+0.13}_{-0.12}$ \\ [4pt]
$\alpha$ & & $-4.958^{+0.018}_{-0.022}$  & $4.930^{+0.016}_{-0.016}$ &$4.948^{+0.010}_{-0.038}$   &  $-4.956^{+0.022}_{-0.011}$\\ [4pt]
$s$ & & $0.468^{+0.013}_{-0.020}$  &$0.456^{+0.016}_{-0.021}$ & $2.018^{+0.011}_{-0.004}$ & $1.998^{+0.015}_{-0.015}$	\\ [4pt]
$q$ & & $0.505^{+0.030}_{-0.058}$  & $0.575^{+0.035}_{-0.044}$ & $0.255^{+0.010}_{-0.010}$ & $0.223^{+0.016}_{-0.001}$ \\ [4pt]
$\pi_{\mathrm{E},\mathrm{N}}$ & & $0.187^{+0.018}_{-0.039}$ &$0.102^{+0.030}_{-0.033}$ &$0.133^{+0.054}_{-0.101}$ &$0.134^{+0.032}_{-0.019}$  \\ [4pt]
$\pi_{\mathrm{E},\mathrm{E}}$ & & $-0.077^{+0.024}_{-0.023}$  & $-0.191^{+0.037}_{-0.024}$ &$-0.270^{+0.030}_{-0.021}$&$-0.164^{+0.032}_{-0.014}$   \\ [4pt]
\hline    
$\chi^2$ & &  485.5  & 487.0 & 486.0 & 487.0 \\   
\end{tabular}
\end{table*}

\begin{table*}
\centering
\small
\caption{Parameters of the four models from modeling of the light curves using the Markov chain. The parameters are obtained by combining ground-based and Gaia data.} 
\label{tab:21blxmod}
\begin{tabular}{cccccc}
\hline
Parameters & (Unit) & C+ & C- & W+ & W-\\
\hline
$t_\mathrm{E}$ & days &  $148.3^{+8.7}_{-7.0}$& $ 134.1^{+18.9}_{-14.1}$ & $103.0^{+6.8}_{-7.3}$ & $116.7^{+10.3}_{-18.7}$ \\  [4pt] 
$t_0$ & HJD-2450000 &  $9300.29^{+0.20}_{-0.51}$ & $9299.66^{+0.16}_{-0.34}$ & $9297.38^{+0.07}_{-0.35}$ & $9297.96^{+0.08}_{-0.08}$ \\  [4pt]
$u_0$ & & $0.0888^{+0.0017}_{-0.0064}$ & $-0.1032^{+0.0127}_{-0.0072}$ &  $0.0323^{+0.0030}_{-0.0030}$ & $-0.0070^{+0.0027}_{-0.0006}$ \\ [4pt]
$\rho_{*}$ & $10^{-3}$&  $1.86 ^{+0.15}_{-0.14}$ & $2.13 ^{+0.24}_{-0.30}$ & $2.67 ^{+0.23}_{-0.21}$ & $2.41 ^{+0.21}_{-0.36}$ \\ [4pt]
$\alpha$ & & $-4.875^{+0.022}_{-0.018}$ & $4.830^{+0.007}_{-0.031}$ &  $4.970^{+0.014}_{-0.003}$ & $-5.030^{+0.080}_{-0.050}$ \\ [4pt]
$s$ & &  $0.432^{+0.006}_{-0.014}$ & $0.464^{+0.017}_{-0.029}$ & $2.013^{+0.010}_{-0.001}$& 	$1.989^{+0.027}_{-0.027}$ \\ [4pt]
$q$ & & $0.555^{+0.067}_{-0.043}$  & $0.491^{+0.023}_{-0.078}$ &  $0.233^{+0.013}_{-0.009}$ & $0.210^{+0.017}_{-0.016}$ \\ [4pt]
$\pi_{\mathrm{E},\mathrm{N}}$ & & $0.134^{+0.018}_{-0.033}$ & $0.088^{+0.056}_{-0.007}$ & $0.233^{+0.022}_{-0.059} $& $0.185^{+0.028}_{-0.062}$ \\ [4pt]
$\pi_{\mathrm{E},\mathrm{E}}$ & &  $0.018^{+0.014}_{-0.035}$ & $ -0.073^{+0.034}_{-0.038}$ &$ -0.202^{+0.024}_{-0.031}$ & $-0.118^{+0.035}_{-0.065}$  \\ [4pt]
$\gamma_{\parallel}$ & year$^{-1}$ & $-1.32^{+0.47}_{-0.15}$ & $-1.76^{+0.66}_{-0.44}$ &  $0.10^{+0.33}_{-0.07}$ & $0.12^{+0.47}_{-0.44}$  \\ [4pt]

$\gamma_{\perp}$ & year$^{-1}$& $-0.16^{+0.29}_{-0.33}$ & $1.77^{+0.58}_{-0.11}$ &  $-0.15^{+0.22}_{-0.07}$ & $1.34^{+0.66}_{-1.02}$  \\ [4pt]
$\gamma_{z}$ & year$^{-1}$ & 
 $<1.92$ &
 $<2.53$ &
 $<1.11$ &
$<2.65$ \\[4pt]
\hline    
$\chi^2$ & &  477.0  & 477.9 & 482.5 & 485.7 \\   
\end{tabular}
\end{table*}


\begin{table*}
\centering
\small
\caption{Magnitudes of the baseline and blend, indicated with the subscript $\mathrm{S}$ and $\mathrm{b}$, respectively, of the four models for each telescope.} 
\label{tab:gd21blxmag}
\begin{tabular}{cccccc}
\hline
   &  Unit  &  C+ & C- & W+ & W-\\
   \hline
$G_\mathrm{S}$ & mag & $19.557 \pm 0.062$ & $19.367 \pm 0.137$ & $18.375 \pm 0.079$ & $18.476 \pm 0.128$ \\
$G_\mathrm{b}$ & mag & $18.200 \pm 0.018$ & $18.261 \pm 0.049$ & $19.101 \pm 0.155$ & $18.927 \pm 0.193$ \\
$g_\mathrm{S}$ & mag & $21.284 \pm 0.065$ & $21.107 \pm 0.143$ & $20.113 \pm 0.073$ & $20.212 \pm 0.135$\\
$g_\mathrm{b}$ & mag & $19.225 \pm 0.010$ & $19.256 \pm 0.027$ & $19.597 \pm 0.045$ & $19.542 \pm 0.073$ \\
$i_\mathrm{S}$ & mag & $19.147 \pm 0.062$ & $18.972 \pm 0.145$ & $17.975 \pm 0.070$ & $18.074 \pm 0.132$\\
$i_\mathrm{b}$ & mag & $17.362 \pm 0.090$ & $17.401 \pm 0.034$ & $17.870 \pm 0.063$ & $17.780 \pm 0.102$ \\
\hline
\end{tabular}
\end{table*}

\section {Constraints on the lens}

From the four models, we derived the source flux and the blend flux listed in Table \ref{tab:gd21blxmag}. The blend is very high for all the models. 
The blend can have multiple origins: it may come from field stars that are very close by, from a second source that is not affected by microlensing, from the lens itself, or from a combination of these cases. In this context, because the lens is a binary system formed by two stars, it is plausible that its flux contributes to the blend or even explains it totally.
In the case of only a partial contribution, we should invoke a second source or a field star to explain the residual blend flux. Although this is still possible, considering that Gaia21blx is a disk event in which the stellar density is lower than in the bulge, the chance alignment of a third object seems relatively unlikely. Therefore, we started with the simplest assumption that the blend is entirely caused by the binary lens itself. This means the smallest number of objects needed to explain the observations, and we verified that the results were fully sound from all physical points of view. Only when we found any inconsistencies would we be pushed to consider a more complicated system with a third system aligned along the same line of sight.
We therefore used the parallax measured by Gaia derived from the Gaia-DR3 release $\pi_\mathrm{Gaia}=(0.45 \pm 0.13$)~mas with a renormalized unit weight error (RUWE) of 1.18. A RUWE with this value means that the parallax measured by Gaia is reliable.If it were higher than 1.4, it would have indicated systematics \citep{GaiaRuwe,Bachelet2022}. \\
At this point, we assumed that the Gaia parallax did not correspond to the source or lens parallax alone, but that it was the flux-weighted average of the parallaxes of the lens and source. Introducing the lens parallax $\pi_\mathrm{L}$ and the source parallax $\pi_\mathrm{S}$, we related these two quantities with $\pi_\mathrm{Gaia}$ using the following equation:
\begin{equation}
\label{eq:pigaia}
\pi_\mathrm{Gaia}=\frac{\pi_\mathrm{S}\, 10^{-0.4\,G_S}+\pi_\mathrm{L}\,10^{-0.4\,G_L}}{10^{-0.4\,G_S}+10^{-0.4\,G_L}}\,,
\end{equation}
where $G_\mathrm{L}=G_\mathrm{b}$.  From now on, we use this equation to replace $\pi_S$ as a function of $\pi_L$.

\subsection{Blend flux constraint}

With the assumption made in Eq.\ref{eq:pigaia}, we have the combined information on the lens and source. We restricted the physical range of the parameters that can be used to explore the region in the $D_\mathrm{L}$--$M$  space where the lens lies. A first way to explore this region is using the mass-luminosity relation (MLR).
Since there are no MLRs in the Gaia band for low-mass stars (\citet{Malkov2022} provided an MLR but stopped at 1.4~$M_{\odot}$)  
we converted the magnitudes into V band, for which MLRs for low-mass stars are available. To do this, we used the relation in \citet{Riello2021}, and we derived the Gaia color $G_\mathrm{BP}-G_\mathrm{RP}$ from the color $g'-i'$ with
\begin{eqnarray}
G_\mathrm{BP}-G_\mathrm{RP} & = & 0.3971+0.777\,(g'-i')-0.4164\,(g'-i')^2+ \nonumber \\
& & \quad +\, 0.008237\,(g'-i')^3\,.
\label{eq:gaiacolor}
\end{eqnarray}

After we obtained the Gaia color from Eq. (\ref{eq:gaiacolor}), we took the equation that relates the Gaia G band with Johnson V band using the Gaia color in order to obtain the V magnitude for the lens,\\
\begin{eqnarray}
G-V & = & -0.02704+0.01424\,(G_\mathrm{BP}-G_\mathrm{RP}) -\nonumber\\
 & & \quad -\,0.2156\,(G_\mathrm{BP}-G_\mathrm{RP})^2+ \nonumber \\
& & \quad +\,0.01426\,(G_\mathrm{BP}-G_\mathrm{RP})^3 \,.
\end{eqnarray}
Including the extinction along the line of sight following the work of \citet{Capitanio2017}, and using the apparent magnitude obtained, we wrote the equation for the absolute magnitude
\begin{equation}
\label{eq:mv}
    M_{\mathrm{V},\mathrm{L}}(M)= V_\mathrm{L}-5 \log(D_\mathrm{L})-A_\mathrm{V}(D_\mathrm{L})\,,
\end{equation}
where the dependence on the lens distance is made explicit. 

At this point, we used the MLR of \citet{Xia2008}. Since the lens is binary, we considered the two components separately, using 
\begin{equation}
\log M_i = \left\{
\begin{array}{l}
0.213-0.0250\,M_{\mathrm{V},i}-0.00275\,M_{\mathrm{V},i}^2\\
\qquad \mbox{for} \quad M_i \in (0.50,1.086)~M_{\odot}\\[1.1ex]
0.982-0.128\,M_{\mathrm{V},i} \\
\qquad \mbox{for} \quad M_i \in [0.28,0.50]~M_{\odot}\\[1.1ex]
4.77-0.714\,M_{\mathrm{V},i}+0.0224\,M_{\mathrm{V},i}^2 \\
\qquad \mbox{for} \quad M_i \in (0.1,0.28)~M_{\odot}
\end{array}
\right.
\end{equation}
and in the end, we summed over the fluxes of the two lenses.

The mass ratio $q$ is fixed by the microlensing modeling, and the total mass is obtained requiring that the total magnitude of the system $M_\mathrm{V}$ matches $M_{\mathrm{V},\mathrm{L}}$ as derived from Eq. \ref{eq:mv}. In this way, we derived the mass as a function of the distance. \\

\subsection{Constraint on the finite source effect}

From the source angular size, we obtained another constraint. Following  \citet{Boyajian2014}, we derived $\theta_*$  using the color $g'-i'$ to derive the zero-magnitude angular diameter $\theta_{m_{\lambda}=0}$, which represents the angular diameter of a star when it is at a distance at which its apparent magnitude equals zero, expressed by the polynomial \citep{Boyajian2014}
\begin{equation}
    \log \theta_{m_{g'}=0}=0.692+0.543\,(g'-i')-0.021\,(g'-i')^2
\end{equation}

The source angular diameter is obtained as \citep{Boyajian2014, Barnes1978}
\begin{equation}
\log \,\theta_*=\log \, \theta_{m_{\lambda}=0}-0.2\,m_{\lambda}\,,
\end{equation}
where $m_{\lambda}$ is the apparent magnitude of a star in a certain filter $\lambda$. In this case, we used the value of $g'_\mathrm{S}$ for our source, as found in Table \ref{tab:gd21blxmag} .\\
In this way, with both $\theta_*$ and $\rho_*$, we were able to derive $\theta_E$ because $\theta_E=\theta_*/\rho_*$. The Einstein angle combined with the Gaia parallax (we recall that $\pi_S$ is obtained by Eq. \ref{eq:pigaia}) gives us another relation of the mass and distance, which is explained as
\begin{equation}
M=\frac{\theta_\mathrm{E}^2}{\kappa\, (\pi_\mathrm{L}-\pi_\mathrm{S})}\,.
\label{eq:massfinite}
\end{equation}

\subsection{Constraint on the microlensing parallax}

We obtained a third constraint using the microlensing parameters, in particular, the parallax obtained from the Markov chain $\pi_E$. Combined with Gaia parallax, this gives us
\begin{equation}
M=\frac{\pi_\mathrm{L}-\pi_\mathrm{S}}{\kappa\,\pi_\mathrm{E}^2}\,.
\label{eq:massparallax}
\end{equation}

\subsection{Combination of the three constraints}

\begin{figure}[!h]
\centering
\includegraphics[width=0.50\textwidth]{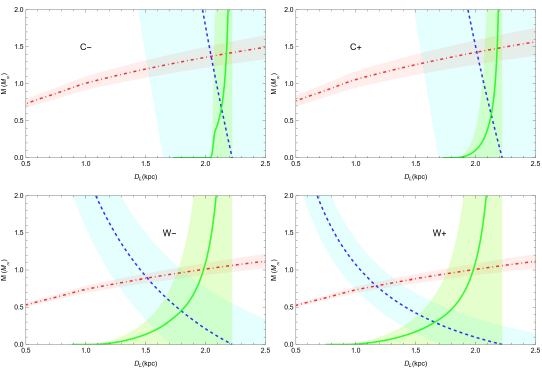}
    \caption{$M$--$D_\mathrm{L}$ region for all the models. The blue stripe with the dashed line represents the $M$--$D_\mathrm{L}$ relation obtained by microlensing parallax ($\pi_\mathrm{E}$). The red stripe with the dot-dashed line shows the constraint from the lens flux, for which the MLRs for low-mass stars are used. The green stripe with the continuous line exhibits the constraint obtained from the finite-source effects \citep{Boyajian2014} and the angular radius of the source. All the stripes have a range of $1 \sigma$} 
    
    \label{fig:gd21blxdolm}
\end{figure}

The three constraints appear as three colored bands whose widths track the uncertainties at 1 $\sigma$, as shown in Fig. \ref{fig:gd21blxdolm}. The parallax error is 30\%, making it the weakest constraint. The most restrictive bound comes from the MLR, while the constraint of finite-source effects becomes an upper bound for the lens distance when $D_L$ and $D_S$ tend to coincide. A compatible solution is found when all three allowed regions overlap.  
This occurs for the two close cases, C+ and C-, while the wide case W- is slightly disfavored, because the overlap region within the allowed stripes at 1 $\sigma$ is small. The W+ is instead excluded at over 5 $\sigma$. We also note that the wide models are also slightly disfavored by the $\chi^2$, as evident in Table \ref{tab:21blxmod}. In Table \ref{tab:g21blxphy}, we report the allowed range for the mass and lens distance by identifying the region in which the three constraints overlap for the best model C+. The C- model gives us a similar result with a $D_\mathrm{L}=2.17 \pm 0.06$ kpc and $M=1.41 \pm 0.15 \, M_{\odot}$ composed oftwo lenses of $M_1=0.94 \pm 0.15 \, M_{\odot}$ and $M_2=0.46 \pm 0.06 \, M_{\odot}$.  The small overlap region for the W- model corresponds to a binary lens of $\sim 0.9 \, M_{\odot}$ that lies at a distance of $\sim 1.8$ kpc, composed of a K star and an M dwarf (see Table \ref{tab:g21blxphyw}). From the combination of the three constraints, we are also able to calculate the Einstein angle $\theta_E=0.69 \pm 0.50$ mas and the projected separation  $a_{\perp}=0.65 \pm 0.47 \, au$. For the W- model, we have $\theta_E=1.16 \pm 0.10$ mas and $a_{\perp}=4.15 \pm 0.36 \, au$ \\

\begin{table*}
    \centering
    \caption{Physical parameters of the lens and source for the best model C+}. 
    \label{tab:g21blxphy}
    \begin{tabular}{ccccccc}
    \hline
   &  Distance (kpc)  & Mass ($M_{\odot}$) & $g'$ & $i'$ & G (Gaia) & Spectral class\\
   \hline
    Total lens & $2.18 \pm 0.07$ & $1.48 \pm 0.17$ & $19.225 \pm 0.010$ & $17.362 \pm 0.090$ & $18.200 \pm 0.018$ & -  \\
    Primary lens & $2.18\pm 0.07$ & $0.95 \pm 0.12$  & $20.023\pm 0.530$ & $17.697\pm 0.530$ & $18.016\pm 0.749$ & G star  \\
   Secondary lens& $2.18\pm 0.07$ & $0.53 \pm 0.07$  & $25.000\pm 0.514$ & $20.601 \pm 0.514$ & $20.411\pm 0.727$ & K star \\
   Source &  $2.38 \pm 0.31$ &$1.10 \pm 0.18$ & $21.284 \pm 0.065$ &  $19.147 \pm 0.062$  & $19.557 \pm 0.062$ & Subgiant F star   \\
\hline
\end{tabular}
\end{table*}

\begin{table*}
    \centering
    \caption{Physical parameters of the lens and source for the alternative model W-}. 
    \label{tab:g21blxphyw}
    \begin{tabular}{ccccccc}
    \hline
   &  Distance (kpc)  & Mass ($M_{\odot}$) & $g'$ & $i'$ & G (Gaia) & Spectral class\\
   \hline
    Total lens & $1.79 \pm 0.07$ & $0.94 \pm 0.09$ & $19.542 \pm 0.073$ & $17.780 \pm 0.102$ & $18.927 \pm 0.193$ & -  \\
    Primary lens & $1.79\pm 0.07$ & $0.78 \pm 0.08$  & $21.312\pm 0.653$ & $18.306\pm 0.653$ & $18.469\pm 0.924$ & K star  \\
   Secondary lens& $1.79\pm 0.07$ & $0.16 \pm 0.02$  & $29.551\pm 0.675$ & $24.401 \pm 0.675$ & $24.575\pm 0.954$ & M star \\
   Source &  $2.63 \pm 0.10$ &$0.96 \pm 0.09$ & $20.212 \pm 0.135$ &  $18.074 \pm 0.132$  & $18.476 \pm 0.128$ & G star   \\
\hline
\end{tabular}
\end{table*}

\begin{figure*}[!h]
\centering
\includegraphics[width=0.65\textwidth]{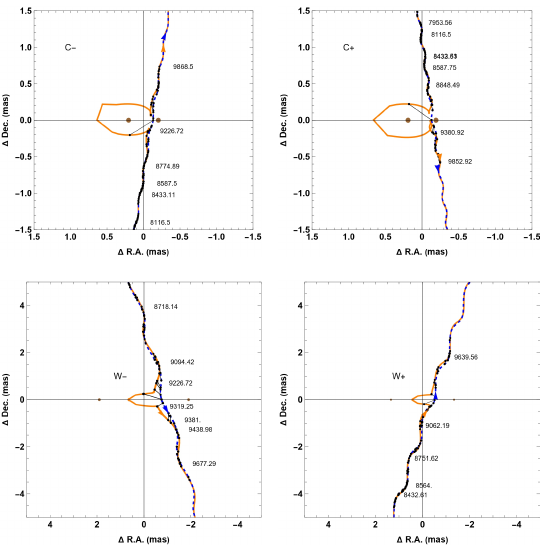}
    \caption{Astrometry simulation for all the models. The blue dashed curve is the source trajectory in a frame in which the lenses are fixed. The orange curve shows the photo-centroid trajectory. The two brown disks are the two lenses and the black points represent the Gaia data points. For all the models the lens flux considered is obtained by the overlap of the two constraints of finite source effects and blend flux.}
    \label{fig:gd21blxas}
\end{figure*}

A further consistency check of our assumption on Gaia parallax is given by astrometry. Using the latest version of VBBinaryLensing of \citet{Bozza2021} we were able to simulate the astrometric microlensing of the event, as shown in Fig. \ref{fig:gd21blxas}. The baseline of Gaia data is very long compared to the microlensing duration. Together with the absence of Gaia points during the magnified section of the light curve, this strengthens the hypothesis that the parallax measured by Gaia is largely unaffected by microlensing and can be considered the flux-weighted average of the parallaxes of the lens and source.

\begin{figure}[!h]
\centering
\includegraphics[width=0.48\textwidth]{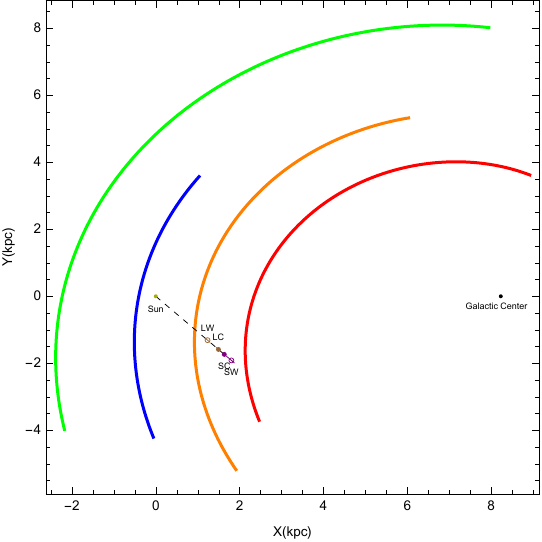}
    \caption{Configuration of Gaia21blx in the Milky Way. The colored spirals represent (from left to right) the Perseus arm (green), the Local arm (blue), the Sagittarius arm (orange) and the Centaurus arm (Red). In the figure, the Galactic center is represented by the black disk, while the lens and source for the C+ model are represented by the brown and purple disks, respectively. For the W- model, empty circles have been used (same colors as in the previous case). The Sun is the darker yellow point. $X$ and $Y$ are the cartesian coordinates in the Galactic plane, centered at the Sun, with the $X$ axis positive toward the Galactic center and the $Y$ axes directed along the rotational curve. The configuration is computed following the work of \citet{Castro2021}.}
    \label{fig:gal}
\end{figure}

Using IAC-STAR \citep{Aparicio2004} with the stellar evolution library of \cite{Girardi2000}, we extrapolated the magnitude in the Johnson Cousins bands. In this way, using the conversion from Johnson to SDSS band \cite{Davenport2006}, we were able to obtain the g' and i' band for each lens.  Using the formulas present in \cite{Riello2021} backwards, the magnitude was also obtained in the Gaia band, as reported in Table \ref{tab:g21blxphy}. We conclude that the source lies at a distance of about 2.4 kpc and probably is an F subgiant star. The lens and source are both located between the Sagittarius and the Centaurus arms (see Fig. \ref{fig:gal}). For the wide solution, we obtain a G star source located in the Centaurus arm at 2.6 kpc from the Sun.

\subsection{Limb Darkening}

We included the limb darkening of the source brightness profile in our models, because the finite-source effects play a significant role. To ensure an accurate estimation of the linear limb-darkening coefficients for each telescope, we proceeded as follows. We used IAC-Star \citep{Aparicio2004} with the stellar evolution library of \cite{Girardi2000} and started from the absolute magnitude in V band obtained for the source. We simulated a stellar population with the solar metallicity, from which we obtained $\log\, g= 4.2_{-0.2}^{+0.3}$ and $T_\mathrm{eff}=6100_{-350}^{+400}~\mathrm{K}$. Using these values, we derived the linear limb-darkening coefficients from \citet{VanHamme1993} for each telescope listed in Table~\ref{tab:data}. The limb darkening does not affect the final result, which we explicitly verified by repeating the analysis with the values at the extremes of the intervals. This is reinforced because the peak lacks enough points to be sensitive to fine details like this.

\subsection{Possible source contamination}

The assumption that the blending flux is generated entirely by the lens does not mean that it is the only plausible hypothesis. We attempted to quantify the probabilities for alternative explanations of the blending flux. First, we considered the hypothesis of a third object that is not associated with either the lens or the source, and that is therefore a field star that affects the blend flux by more than 10\% at least. Using the latest updated version of the Besançon model \citep{Lagarde2017, Lagarde2019}, we simulated a stellar population in the vicinity of the target, that covered an area of 0.2 square degrees. Taking into account that the minimum separation that Gaia can detect is 0.05 square arcseconds \citep{deBruijne2015}, we calculated the probability of a random alignment of a third object. we obtained a value of 0.001\% that means that the hypothesis that there is a third object is therefore very unlikely. In this case, the blending flux would be an upper limit on the flux of the lens and hence on the mass. Considering both the best models obtained from microlensing and the fact that the RUWE obtained from GaiaDR3 is 1.14, we can exclude the presence of a third lens or a nearby secondary source. Regarding the possibility that the contaminant is a very wide companion of the source and considering that for F, G, K type stars, the probability that they are part of binary or multiple systems is 46\% \citep{Raghavan2010}. 
Based on the previously calculated $\theta_E=0.69 \pm 0.50$ mas, in order for the source to have a companion that is not detected by Gaia nor in the microlensing light curve, it must be in a region that extends from the minimum resolution of Gaia of 230 mas and $1/4{\theta_E}=0.36$ mas. Following the work of \citet{Fukui2015}, we calculated the probability that a companion source lies in this range. First, using the minimum resolution of Gaia and the maximum angular that can be detected by microlensing, and always assuming a source distance of 2.38 kpc, we obtained a separation range between the two companions of 1-550 au. From the Besançon model, always considering a contribution of  the secondary source to the blend flux of 10\%, we obtained a mass of the latter equal to $0.70 M_{\odot}$. Applying Kepler's third law, we obtained the period range of this hypothetical binary system, which must be $2.3 < Log (P) < 6.5$. Considering the probability of a binary or multiple system alone from \citet{Raghavan2010} of 46\%, and using, as we did throughout, the period log-normal distribution with mean $P=5.03$ and standard deviation $\sigma_{log P}=2.28$, we derived the fraction of binaries in the range of periods calculated earlier based on this, which is 62\%. Finally, considering the 46\% probability of a binary or multiple system for F-G-K stars, we obtained a total probability of 28\%. Following the same procedure, we calculated the probability that a third lens acts as a blending lens that has no effect on microlensing, otherwise, there would have been further peaks in the light curve. Always considering a contribution of $10\%$ of the total blending flux, following the same approach as for a hypothetical second source, we obtained that the probability of having a third lens is $9\%$.
We repeated the same approach for the wide models and calculated that the probability of a secondary source that is not affected by microlensing is equal to 30\%. The probability of a third star in the lens system is 10 \%.

We conclude that all three alternative hypotheses are mildly or strongly disfavored. The most competitive alternative to our proposed interpretation is a companion to the source. Possible observations that confirm whether our main conclusion or one of the enumerated alternatives is correct may include high-resolution spectroscopic studies that may confirm the presence of one source and two lenses, as described in our model. The relative proper motion is quite low (see Sect. \ref{Sec kinematics}), which discourages high-resolution imaging for the separation of lens and source.

\section{Lens and source kinematics} \label{Sec kinematics}

With the precise determination of the lens mass and distance both according to model C+ and to model W-, we are in the position tu fully describe the kinematics of the lens and source systems. We describe all details for model C+ and quote the results for W- at the end of the section.
For model C+, the relative lens-source proper motion in the geocentric reference frame is
\begin{equation}
    \mu_\mathrm{rel}=\frac{\kappa\,M\,\pi_\mathrm{E}}{t_\mathrm{E}} = (1.69 \pm 1.22)~\mathrm{mas/yr}\,.
\end{equation}

Using the eastern and northern components of the parallax vector, we obtained the components of the proper motion in the geocentric frame as
\begin{equation}
\begin{split}
    \vec{\mu}_{\mathrm{rel},\mathrm{geo}} 
     &=\frac{\mu_\mathrm{rel}}{\pi_\mathrm{E}}\,(\pi_{\mathrm{E},\mathrm{E}}\,\hat{\vec{E}}+\pi_{\mathrm{E},\mathrm{N}}\,\hat{\vec{N}}) \\
    &= [(0.23 \pm 0.85)\,\hat{\vec{E}}+ (1.67 \pm 0.88)\,\hat{\vec{N}}]~\mathrm{mas}/\mathrm{yr}\,,
\end{split}
\end{equation}
where $\hat{\vec{E}}$ and 
$\hat{\vec{N}}$ are the unit vectors in the plane orthogonal to the line of sight: $\hat{\vec{N}}$ is tangent to the celestial meridian pointing to the north while $\hat{\vec{E}}$ is tangent to the celestial parallel pointing to the east.
It is simple to convert them into the heliocentric reference frame by using the velocity vectors of the Earth at time $t_0$ projected orthogonally to the line of sight,
\begin{eqnarray}
    \vec{\mu}_{\mathrm{rel},\mathrm{hel}} & = &\vec\mu_{\mathrm{rel},\mathrm{geo}} + \vec{v}_{\oplus}\,\frac{\pi_\mathrm{rel}}{\mathrm{au}} \nonumber \\
    & = &[(0.41 \pm 0.19)\,\hat{\vec{E}}+(1.75 \pm 0.12)\,\hat{\vec{N}}]~\mathrm{mas}/\mathrm{yr}\,.
    \label{eq:muhel}
\end{eqnarray}

Gaia DR3 release also reports proper motion components for Gaia21blx \citep{Riello2021}. As for the parallax (see Eq.\ref{eq:pigaia}), we considered the proper motion measured by Gaia as the flux-weighted average of the lens and source proper motions,
\begin{equation}
    \begin{split}
    \vec{\mu}_{\mathrm{Gaia}}&=\frac{\vec{\mu}_\mathrm{S}\,10^{-0.4\,G_\mathrm{S}}+\vec{\mu}_\mathrm{L}\,10^{-0.4\,G_\mathrm{L}}}{10^{-0.4\,G_\mathrm{S}}+10^{-0.4\,G_\mathrm{L}}} \\
    &= [(-6.81 \pm 0.11)\,\hat{\vec{E}} + (-3.24 \pm 0.12)\,\hat{\vec{N}}]~\mathrm{mas}/\mathrm{yr}
    \label{eq:mugaia}
    \end{split}
\end{equation}

Noting that the relative proper motion is simply the difference between the lens and source proper motions ($ \vec{\mu}_{\mathrm{rel},\mathrm{hel}} \equiv \vec{\mu}_\mathrm{L} - \vec{\mu}_\mathrm{S}$), Eqs. (\ref{eq:muhel}) and (\ref{eq:mugaia}) represent two independent linear constraints for $\vec{\mu}_\mathrm{L}$ and $\vec{\mu}_\mathrm{S}$, which can easily be solved to obtain
\begin{equation}
\begin{split}
    \vec{\mu}_\mathrm{L}&=[(-6.73\pm 0.48)\,\hat{\vec{E}} + (-2.87 \pm 0.08)\,\hat{\vec{N}}]~\mathrm{mas}/\mathrm{yr}\,, \\
    \vec{\mu}_\mathrm{S}&=[(-7.14\pm 0.48)\,\hat{\vec{E}} + (-4.55\pm 0.08)\,\hat{\vec{N}}]~\mathrm{mas}/\mathrm{yr}\,.
\end{split}
\end{equation}

We then applied a $26.84$° rotation to these vectors to determine their components in relation to the Galactic frame as
\begin{equation}
    \begin{split}
        \vec{\mu}_\mathrm{L} &= [(-7.30\pm 0.48)\,\hat{\vec{l}} + (0.48 \pm 0.08)\,\hat{\vec{b}}]~\mathrm{mas}/\mathrm{yr}\,, \\
        \vec{\mu}_\mathrm{S} &= [(-8.42\pm 0.48)\,\hat{\vec{l}}  + (-0.84 \pm 0.08)\,\hat{\vec{b}}]~\mathrm{mas}/\mathrm{yr}\,,
    \end{split}
\end{equation}
where we introduced unit vectors pointing in the direction of increasing Galactic longitude $\hat{\vec{l}}$ and increasing Galactic latitude $\hat{\vec{b}}$.\\
We know the distances $D_L$ and $D_S$ and therefore switched from the proper motion to the heliocentric velocity components
\begin{equation}
    \begin{split}
        \vec{v}_{\mathrm{L},\mathrm{hel}}&= [(-69.35 \pm,4.96)\,\hat{\vec{l}} +(-29.68\pm 0.81)\,\hat{\vec{b}}]~\mathrm{km}/\mathrm{s}\,, \\
        \vec{v}_{\mathrm{S},\mathrm{hel}}&= [(-80.62\pm 5.41)\,\hat{\vec{l}} + (-51.35\pm 0.89)\,\hat{\vec{b}}]~\mathrm{km}/\mathrm{s}\,.
    \end{split}
\end{equation}

It is interesting to calculate the peculiar velocities of the lens and source with respect to their local standard of rest in order to assign these objects to definite kinematic components of the Galaxy. In order to do this, we converted the heliocentric velocities into Galactocentric velocities and then subtracted the average rotation of the disk at the position of the lens and source. A fully 3D reconstruction of the peculiar motion is not possible because the radial velocities of the lens and source are lacking. However, as long as we just have to add or subtract vectors, we have all the information needed to calculate the components along the directions $\hat{\vec{l}}$  and $\hat{\vec{b}}$ orthogonal to the line of sight, which already contain sufficiently interesting information, as we show below.

The galactocentric velocities were obtained by adding the solar velocity, including the local standard of rest $\boldsymbol{\Theta}_\sun$ and the peculiar velocity $u_\sun$,
\begin{equation}
\begin{split}
       \vec{v}_{\mathrm{L},\mathrm{gal}}&=\vec{v}_{\mathrm{L},\mathrm{hel}}+\vec{\Theta}_{\odot} + \vec{u}_\odot \,,\\
       \vec{v}_{\mathrm{S},\mathrm{gal}}&=\vec{v}_{\mathrm{S},\mathrm{hel}}+\vec{\Theta_{\odot}} + \vec{u}_\odot\,.
\end{split}
\end{equation}

We took the circular orbital speed at the Sun $\Theta_0= (236 \pm 7)~\mathrm{km}/\mathrm{s}$
and the peculiar velocity $\vec{u}_{\odot}=(18.5\,\hat{\vec{l}} + 7.1\,\hat{\vec{b}})~\mathrm{km}/\mathrm{s}$ \citep{Reid2019}.

The average rotation of the disk is described by a rotation curve $\Theta(r)$ as a function of the galactocentric distance $r$. We assumed for simplicity a flat rotation curve with $\Theta(r)=\Theta_\odot$. The peculiar velocities are then
\begin{equation}
\begin{split}
       \vec{u}_\mathrm{L}&=\vec{v}_{\mathrm{L},\mathrm{gal}}-\vec{\Theta}(r_\mathrm{L})\,, \\
       \vec{u}_\mathrm{S}&=\vec{v}_{\mathrm{S},\mathrm{gal}}-\vec{\Theta}(r_\mathrm{S})\,.
\end{split}
\end{equation}

The vector $\vec{\Theta}(r)$ is obviously tangent to a circle centered on the Galactic center with radius $r$. In our case, we have $r_\mathrm{L}=(6.81 \pm 0.12)~\mathrm{kpc}$ and $r_\mathrm{S}=(6.70 \pm 0.20)~\mathrm{kpc}$. Finally, we obtained
\begin{equation}
    \begin{split}
        \vec{u}_\mathrm{L}&= [(-17.30\pm 3.92)\,\hat{\Vec{l}} +(13.45 \pm 0.85)\,\hat{\Vec{b}}] \, \mathrm{km}/\mathrm{s} \,,\\
        \vec{u}_\mathrm{S}&= [(-32.03\pm 6.60)\,\hat{\Vec{l}} + (-0.74\pm 0.92)\,\hat{\Vec{b}}] \, \mathrm{km}/\mathrm{s}\,.
    \end{split}
\end{equation}
 Following the same approach for the W- model, we obtained $\vec{u}_\mathrm{L}=[(-16.06\pm 4.32)\,\hat{\Vec{l}} +(-6.04 \pm 1.25)\,\hat{\Vec{b}}]$ and $\vec{u}_\mathrm{S}=[(-46.41\pm 3.93)\,\hat{\Vec{l}} + (26.70\pm 1.82)\,\hat{\Vec{b}}]$.
We next discuss the kinematics of the source and the lens using their peculiar velocities with respect to the average disk rotation.
For model C+, the components along the Galactic latitude are relatively small, which confirms that the two objects share kinematic properties with the Galactic disk. The negative sign in the longitudinal component indicates that both objects seem to be moving away from the Galactic center and/or are at a lower velocity with respect to the assumed average rotation $\theta(r)$. In any case, the magnitude of this component is not too large, considering the uncertainties in Galactic rotation. From the low values of the components of the peculiar velocities, we conclude that both the lens and the source both follow the average rotation of the Galactic disk very closely and can thus be assigned to the thin-disk component from a kinematic point of view. The results obtained for the wide model W- also describe stars that seem to be moving away from the Galactic center. The different signs for the components along the Galactic latitude are due to the different direction of the parallax that occurs in the two cases. Even in this configuration, however, we obtain objects with typical properties of Galactic disk stars. The higher peculiar velocity of the source star may be consistent with a more evolved object, as suggested by its color-magnitude position above the main sequence. 

\section{Conclusions}
We analyzed the microlensing event Gaia21blx, which was discovered by Gaia and was densely observed by the OMEGA collaboration. We modeled four degenerate solutions with a good measure of the finite-source effect and a suboptimal measure of the parallax. 
The microlensing models were complemented by the information provided by Gaia. In particular, we assumed that the parallax and the proper motion measured by Gaia are the flux-weighted averages of the parallaxes and proper motion of the lens and source. Combining this assumption with constraints on the angular radius, the mass-luminosity relations, and the microlensing parallax, we examined allowed lens masses and distances for our four models. 

We find a better consistency for the two close-binary models, which correspond to a binary lens system located at 2.2 kpc. In this case, the lens is composed of a G star with 0.9 $M_{\odot}$ and a K star with 0.5 $M_{\odot}$. We infer that the source is most likely a subgiant F star located at 2.4 kpc. The lens and the source would both belong to the thin Galactic disk and follow the average rotation rate very closely, lying in a region between the Sagittarius and the Centaurus arms. However, an alternative wide-binary model remains, although it is slightly disfavored by $\chi^2$ and by the parallax constraint. In this case, the binary lens consists of a K-type star and an M dwarf located at 1.8 kpc, with the source at 2.6 kpc. In this case, the lens would be part of the stellar population of the Sagittarius arm, while the source would belong to the Centaurus arm.

Our conclusions can be confirmed by spectroscopic observations that might highlight the lines from the source and from at least the primary lens (since the secondary lens is probably a K star it will be difficult to obtain its spectral lines), which contributes most of the blending. The novel approach introduced here for the analysis of Gaia21blx can be applied to similar events in which the blending is very high and fully attributed to the lens. Our assumption has led to a fully consistent physical picture from the point of view of photometry, astrometry and Galactic kinematics. The chances are very low that random stars in the Galactic disk that are not involved in the microlensing event contribute to the blending, while possible contamination by wide-binary companions of the source cannot be excluded based on available data. This possibility could be tested by spectroscopic observations (with the ESO ESPRESSO \citep{Pepe2021}), while high-resolution follow-up to separate the source and the lens is discouraged by the low relative proper motion.

\begin{acknowledgements}
 This work has made use of the IAC-STAR Synthetic CMD
computation code. IAC-STAR is supported and maintained by
the IT department of the Instituto de Astrofísica de Canarias.\\
EB gratefully acknowledge support from NASA grant 80NSSC19K0291. {\bf This research has made use of the NASA Exoplanet Archive, which is operated by the California Institute of Technology, under contract with the National Aeronautics and Space Administration under the Exoplanet Exploration Program.}
This work has made use of data from the European Space Agency (ESA) mission
{\it Gaia} (\url{https://www.cosmos.esa.int/gaia}), processed by the {\it Gaia}
Data Processing and Analysis Consortium (DPAC,
\url{https://www.cosmos.esa.int/web/gaia/dpac/consortium}) and 
 the Photometric Science Alerts Team (http://gsaweb.ast.cam.ac.uk/alerts).
Funding for the DPAC
has been provided by national institutions, in particular the institutions
participating in the {\it Gaia} Multilateral Agreement. This work makes use of observations from the Las Cumbres Observatory global telescope network. YT acknowledges the support of DFG priority program SPP 1992 “Exploring the Diversity of Extrasolar Planets” (TS 356/3-1). RFJ acknowledges support for this project provided by ANID's Millennium Science Initiative through grant ICN12\textunderscore 009, awarded to the Millennium Institute of Astrophysics (MAS), and by ANID's Basal project FB210003.
This work is supported by Polish MNiSW grant DIR/WK/2018/12 and  European Union's Horizon 2020 research and innovation program under grant agreement No. 101004719 (OPTICON-RadioNet Pilot, ORP).
We also acknowledge support by the italian PRIN 2022J4H55R - Detection of Earth-like ExoPlanets, CUP D53D23002590006. N.P.’s work was supported by Fundação para a Ciência e a Tecnologia (FCT) through the research grants UIDB/04434/2020 and UIDP/04434/2020. We acknowledge support from the Novo Nordisk Foundation Interdisciplinary Synergy Program grant no. NNF19OC0057374.

\end{acknowledgements}

\bibliographystyle{aasjournal}  
\bibliography{gaia21blx_corr.bib}

%
%



\end{document}